\documentclass[sn-mathphys,Numbered]{sn-jnl}

\usepackage{graphicx}%
\usepackage{multirow}%
\usepackage{amsmath,amssymb,amsfonts}%
\usepackage{amsthm}%
\usepackage{mathrsfs}%
\usepackage[title]{appendix}%
\usepackage{xcolor}%
\usepackage{textcomp}%
\usepackage{manyfoot}%
\usepackage{booktabs}%
\usepackage{algorithm}%
\usepackage{algorithmicx}%
\usepackage{algpseudocode}%
\usepackage{listings}%

\usepackage{bm}
\usepackage{color}
\usepackage{here}

\newcommand{\la}{\langle}
\newcommand{\ra}{\rangle}

\theoremstyle{thmstyleone}%
\theoremstyle{thmstyletwo}%

\theoremstyle{thmstylethree}%

\begin{document}

\title[Article Title]{Impact of density inhomogeneity on the critical velocity for vortex shedding in a harmonically trapped  Bose-Einstein condensate}

\author*[1]{\fnm{Haruya} \sur{Kokubo}}\email{kharu\_phys@kindai.ac.jp}

\author[1]{\fnm{Kenichi} \sur{Kasamatsu}}\email{kenichi@phys.kindai.ac.jp}
\equalcont{These authors contributed equally to this work.}

\affil*[1]{\orgdiv{Department of Physics}, \orgname{Kindai University}, \orgaddress{\street{3-4-1 Kowakae}, \city{Higashi Osaka-city}, \postcode{577-8502}, \state{Osaka}, \country{Japan}}}


\abstract{We report on a numerical study of the critical velocity for creation of quantized vortices by a moving Gaussian obstacle in a trapped Bose-Einstein condensate, modeled by the Gross-Pitaevskii equation. 
We pay attention to impact of density inhomogeneity associated with the global inverted-parabolic profile by a trapping potential as well as the local density suppression around the Gaussian obstacle.
When the width of the Gaussian potential is large, the wake dynamics is significantly influenced by the nonuniformity around the obstacle potential. 
The critical velocity, estimated through the time interval between the first and second vortex emission, 
can be explained by the local sound velocity by taking into account the above two contributions.
We also find that the ratio of the critical velocity to the sound velocity at the center of the system is insensitive to the nonlinear coefficient of the Gross-Pitaevskii equation, which supports the universal discussion even in a inhomogeneous trapped condensate under the local density approximation.  

}




\maketitle

\section{INTRODUCTION} \label{intro}
Hydrodynamic instability is one of the key research topics in a fluid dynamics when considering problems of a turbulent transition of a wake flow, which occurs behind an obstacle moving in a fluid \cite{landauTheorySuperfluidityHelium1941}.
In classical hydrodynamics, the characteristics of wake dynamics can be classified by the dimensionless Reynolds number defined as a ratio of an inertial force and a viscous one.

A wake flow has been also studied very well in superfluids, having been realized experimentally in cold atomic-gas Bose-Einstein condensates (BECs) \cite{ramanEvidenceCriticalVelocity1999,onofrioObservationSuperfluidFlow2000,kwonCriticalVelocityVortex2015,kwonObservationArmVortex2016,limVortexsheddingfrequency2022}  or exciton-polariton condensates \cite{amoSuperfluiditypolaritonssemiconductor2009,amoPolaritonSuperfluidsReveal2011,nardinHydrodynamicnucleationquantized2011,lerarioRoomtemperaturesuperfluiditypolariton2017}. 
The wake is achieved in these systems by inducing a condensate flow in the presence of an obstacle potential created by a localized external field. 
There have been lots of theoretical works concerning wake superflows based on the Gross-Pitaevskii (GP) model \cite{frischTransitiondissipationmodel1992,noreSubcriticalDissipationThreeDimensional2000,aftalionDissipativeFlowVortex2003,reevesIdentifyingSuperfluidReynolds2015,Kwak2023-la,staggQuantumAnaloguesClassical2014,jacksonVortexFormationDilute1998,winieckiVortexsheddingdrag2000,Musser2019-xc,Kiehn2022-xk,Kunimi2015-nk,Pinsker2014-el}. 
The fundamental interest in these works is to reveal similarity and difference from the classical counterpart; characteristics in superfluids are mainly due to the presence of the critical velocity of superfluidity and of vortices with quantized circulation. 
These features yield a new platform to study a quantum analogue of the von K\`{a}rm\`{a}n vortex street \cite{kwonObservationArmVortex2016,staggQuantumAnaloguesClassical2014,sasakiEnardArmVortex2010} and the Reynolds number \cite{reevesIdentifyingSuperfluidReynolds2015}. 

To study the wake phenomenon in superfluids, a motion of an obstacle in a static superfluid frame must be above a critical velocity, below which an obstacle motion is frictionless.
However, since the critical velocity is determined by various details, e.g. the shape or the boundary condition, of the obstacle, the systematic determination of the critical velocity is an unresolved issue.
The well-known Landau criterion of superfluidity predicts that the critical velocity in a uniform system is given by the sound velocity $c_s$ by considering the energetic instability for elementary excitations.
When there is a hard wall cylinder in incompressible flow, the local velocity is multiplied by a factor 2 on the lateral side of the cylinder, so that the critical velocity is expected to be $0.5 c_s$.  
In the calculation of the GP model with contributions of compressibility and quantum pressure, the critical velocity is $0.37c_s$ when the size of the cylinder is sufficiently large compared to the healing length \cite{ricaremarkcriticalspeed2001,phamBoundaryLayersEmitted2005}.
In the limit of the vanishing obstacle size, the critical velocity approaches to $c_s$ \cite{ricaremarkcriticalspeed2001,phamBoundaryLayersEmitted2005}.

While the above-mentioned works have been discussed for a uniform system, the atomic-gas BECs are essentially inhomogeneous \cite{ramanEvidenceCriticalVelocity1999,onofrioObservationSuperfluidFlow2000,kwonCriticalVelocityVortex2015,kwonObservationArmVortex2016,limVortexsheddingfrequency2022}.
There are two external contributions that induce inhomogeneity, namely a trap potential and an obstacle potential.
For a BEC trapped in a harmonic potential, the global form of the density profile is an inverted parabola.
In addition, the obstacles are usually modeled by the Gaussian form, which intrinsically involves inhomogeneous distribution of the condensate density around the localized obstacles.  
The previous experimental works \cite{kwonCriticalVelocityVortex2015} clarified that the critical velocity decreases with increasing the width of the Gaussian obstacle.
One possible reason for this decrease could be the appearance of an inhomogeneous region around the Gaussian obstacle.

In a uniform system, the GP equation is usually scaled by the bulk density and the associated healing length, so that the results are free from the coupling constant of the nonlinear term. 
For trapped systems, however, the dependence of the coupling constant is nontrivial because of the additional length scale due to the harmonic trap.
By reducing the nonlinear coefficient, i.e., the particle number or the (repulsive) interaction energy, the global size of the condensate decreases while the healing length increases, the inhomogeneity thus being reinforced. 
It is noticeable that the linear Schr\"{o}dinger limit gives rise to the vanishing critical velocity. 
Thus, it is expected that the inhomogeneity basically plays a role of decreasing the critical velocity. 

In this study, we investigate the critical velocity of the two-dimensional BEC trapped in a harmonic potential for vortex generation by a moving Gaussian obstacle.
To focus on the inhomogeneous effects of vortex generation, we consider the GP model with the nonlinear coefficients smaller than those corresponding to the previous experiments \cite{kwonCriticalVelocityVortex2015}.
Numerical simulations of the GP equation demonstrate that the ratio of the critical velocity to the sound velocity is smaller than previous literatures.
This result suggests that the critical velocity is influenced by the inhomogeneity of the system.
We take into account the local sound velocity due to the trap potential and the Gaussian obstacle to determine which is responsible for the decrease in the critical velocity.

The paper is organized as follows. In Sec.\ref{sec2} we introduce the GP equation with an obstacle potential, discussing the inhomogeneous features of the system caused by the trapping potential and the obstacle potential. In Sec.\ref{sec3} we introduce the method to determine the critical velocity and analyze impact of density inhomogeneity on the critical velocity. Finally, Sec.\ref{sec4} devotes to the conclusions.

\section{FORMULATION}\label{sec2}
Ignoring the inhomogeneity along the axial $z$-direction, we consider a two-dimensional (2D) BEC in the $x$-$y$ space.
Within a mean-field theory, a BEC at low temperatures is described by the macroscopic wave function $\Psi(\bm{r},t)=\sqrt{n(\bm{r},t)}e^{i\theta(\bm{r},t)}$ with the particle density $n$ and the phase $\theta$. 
The dynamics of the wave function obeys the GP equation
\begin{equation}
i\hbar\frac{\partial}{\partial t}\Psi=\left(-\frac{\hbar^2}{2m}\nabla^2+V_{\rm{ext}}+g|\Psi|^2\right)\Psi.
\label{1st_gpe}
\end{equation}
Here, $m$ is the atomic mass, $V_{\rm ext}(\bm{r})$ is the harmonic trap potential and $g$ in the nonlinear term is the coupling constant in the 2D system with the axial contribution being integrated out \cite{reevesIdentifyingSuperfluidReynolds2015,jacksonVortexFormationDilute1998}. 

\subsection{Inhomogeneity by a trapping potential}
First, we confirm the effect by the trapping potential.
The stationary state of the system can be calculated by inserting the form $\Psi(\bm{r},t) = \Phi(\bm{r}) e^{-i \mu t/\hbar}$ into Eq.\eqref{1st_gpe} and by solving the time-independent GP equation
\begin{equation}
\left[-\frac{\hbar^2}{2m}\nabla^2+V_{\rm ext}-\mu+g|\Phi|^2\right]\Phi=0.
\end{equation}
A stationary density profile in a harmonic potential $V_{\rm ext}(\bm{r})=\frac{1}{2}m\omega^2(x^2+y^2)$ is written within the Thomas-Fermi approximation as \cite{baymGroundStatePropertiesMagnetically1996}
\begin{equation}
n_{\rm TF} (\bm{r})=n_0\left(1-\frac{x^2}{R^2_{\rm TF}}-\frac{y^2}{R^2_{\rm TF}}\right).
\end{equation}
Here, the Thomas-Fermi radius is $R_{\rm TF}=\sqrt{{2\mu}/{m\omega^2}}$ and the density at the trap center is $n_0=n(0)={\mu}/{g}$.
From the normalization condition $\int d\bm{r} n_{\rm TF} (\bm{r})=N$, a relationship between the coupling constant $g$ and the chemical potential $\mu$ is given by
\begin{equation}
g=\frac{\pi\mu^2}{mN\omega^2}.
\label{coupling}
\end{equation}
Here, $N$ is a particle number in the 2D system.
We regard the chemical potential $\mu$ as a parameter to represent the coupling constant of the GP equation instead of $g$, which is useful in the following discussion because the relevant energy scales are often compared with $\mu$. We also confine ourselves to the Thomas-Fermi regime $\mu \gg \hbar \omega$ or, equivalently, $R_\text{TF} \gg \xi$ with the healing length $\xi = \hbar/\sqrt{2m\mu}$. 

\subsection{Inhomogeneity by an obstacle potential}
We next consider a 2D trapped BEC with a moving obstacle potential at a constant velocity. 
This system is described by the time-dependent GP equation
\begin{equation}
	i\hbar\frac{\partial}{\partial t}\Psi=\left[-\frac{\hbar^2}{2m}\nabla^2+V_{\rm{ext}}+V_{\rm{obst}}+g|\Psi|^2\right]\Psi .
\end{equation} 
The obstacle potential $V_{\rm obst} (\bm{r},t)$ is given by the Gaussian form as
\begin{equation}
V_{\rm obst}=V_0 {\rm exp}\{-2[(x-x_0-v_{\rm obst}t)^2+y^2]/\sigma^2\},
\label{Vobst}
\end{equation}
which moves at a constant velocity $v_{\rm obst}$ along the $x$ axis.
The parameters $V_0$, $x_0$, and $\sigma$ are the height, the initial position, and the size of the Gaussian potential, respectively.
In this work, the length, time and energy scales are $\sqrt{\hbar/m\omega}\equiv a_h$, $1/\omega\equiv \tau$ and $\hbar \omega$, respectively, by using units of a harmonic potential.
Then, the dimensionless GP equation with a moving Gaussian obstacle is written as
\begin{equation}
i\frac{\partial}{\partial \tilde{t}}\tilde{\Psi}=\left[-\frac{1}{2}\tilde{\nabla}^2+\tilde{V}_{\rm ext}+\tilde{V}_{\rm obst}+\pi\tilde{\mu}^2|\tilde{\Psi}|^2\right]\tilde{\Psi}.
\label{GPE}
\end{equation}
Here, the quantities with tildes are dimensionless and the wave function is scaled as $\tilde{\Psi}=\Psi(\bm{r},t)(a_h/\sqrt{N})$.
The coupling constant $g$ is replaced by $\mu$ through the relation of Eq.\eqref{coupling}; although Eq.\eqref{coupling} is not correct in the presence of $V_\text{obst}$, the resultant correction is negligibly small in our parameter setting as discussed below. 
When the amplitude $V_0$ of the Gaussian obstacle is smaller than the chemical potential $\mu$, the wave function can penetrate into the obstacle's area \cite{kwonCriticalVelocityVortex2015}.
In this work, we focus on the impenetrable regime $V_0 > \mu$, in which the condensate density strongly decays at the boundary of the obstacle and vanishes inside the obstacle. 
The stationary solution with the Gaussian obstacle is obtained by using the imaginary time evolution method for Eq.\eqref{GPE}.
\begin{figure}[ht]
    \centering
    \includegraphics[width=0.7\linewidth]{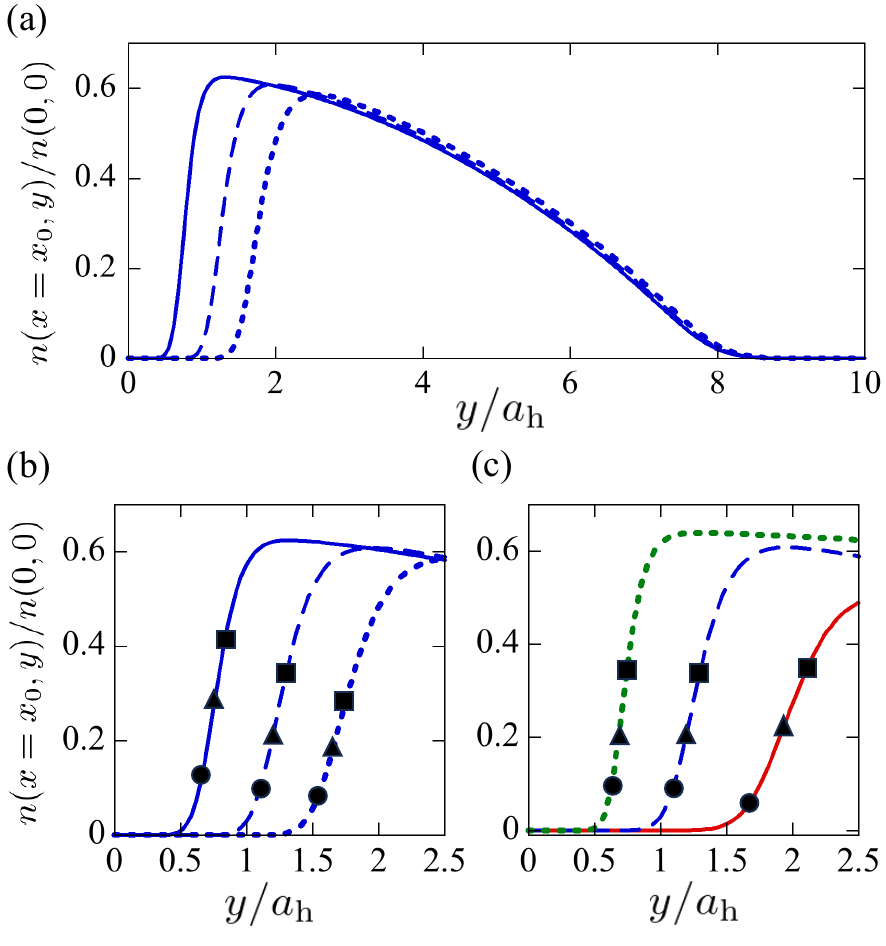}
    \caption{The cross sections of the stationary density $n_0(-0.6R_{\rm TF},y)$ along the $y$-direction passing through the center of the Gaussian obstacle.
(a) The density profiles for $\mu/(\hbar \omega)=50$ and different values of the Gaussian widths $\sigma/\xi=6$ (solid curve), $10$ (dashed curve) and $14$ (dotted curve).
(b) The enlarged plot of (a) around the Gaussian obstacle.
The circles, triangles, and squares correspond to the density at $y=R_{\rm obst}$, $R_{\rm obst} + \xi$, and $R_{\rm obst} + 2 \xi$, respectively, determined in Sec.~\ref{res1}.
(c) A similar plot of (b), but for fixed $\sigma/\xi=10$ and different values of $\mu/ (\hbar \omega) = 20$ (solid curve), 50 (dashed curve), 150 (dotted curve).}
    \label{grad_scale}
\end{figure}
Figure \ref{grad_scale} shows the cross section of the stationary density profile along the $y$-direction across the central position of  the Gaussian obstacle, where we set $x_0 = -0.6R_\text{TF}$ and represent the value of $\sigma$ in units of $\xi$, following the calculation in the next section.
As shown in Fig.\ref{grad_scale}(a), with increasing the Gaussian width $\sigma$ for fixed $\mu$, the size of the density hole, created by Gaussian obstacle, is obviously increased. 
On the other hands, the outer Thomas-Fermi boundary is not influenced so much by the obstacle's size, which supports our statement below Eq.\eqref{GPE}. 
It is noteworthy that, as clearly seen in Fig.\ref{grad_scale}(b), the Gaussian tail becomes larger with increasing $\sigma$, extending the ``recovery region" of the density from zero to the equilibrium Thomas-Fermi value. 
This feature indicates that the density inhomogeneity is more apparent for a larger Gaussian obstacle and has a strong impact to the suppression of the critical velocity as discussed below. 
Also, the recovery region is extended with decreasing the chemical potential $\mu$ even for fixed $\sigma$, as seen in Fig.\ref{grad_scale}(c). 
This is because, the smaller $\mu$ becomes, the longer the healing length $\xi$ is. 
\section{RESULTS}\label{sec3}
We investigate the effects of inhomogeneous density to the wake dynamics and the associated critical velocity by the Gaussian obstacle potential and the trapping potential through the numerical calculation of Eq.\eqref{GPE}.
Since the sound velocity depends on the local density as $c_s(\bm{r}) = \sqrt{g n(\bm{r})/m}$ within the local density approximation, the sound velocity $c_s(\bm{r} = 0) \equiv c_s$ at the center of the system is used as the basis for unifying the scale of the critical velocity of the system.
The initial position of the obstacle is set as $x_0=-0.6 R_{\rm TF}$, and the center of the obstacle moves to the $+x$-direction with the distance $2|x_0|$.
The mesh sizes $\Delta_{x,y}$ are determined by $\Delta_{x,y}=\xi/4$ and the time step is $\Delta t = 5.0\times10^{-6}\tau$.
In this study, the chemical potential $\mu$ and the width $\sigma$ of the Gaussian potential are smaller than those of the experiment \cite{kwonCriticalVelocityVortex2015}.
The value of $V_0$ in Sec.\ref{res1} is taken to be the same value as that of experiment for comparison with the experiment \cite{kwonCriticalVelocityVortex2015}.
\begin{figure}[ht]
	\includegraphics[width=\linewidth]{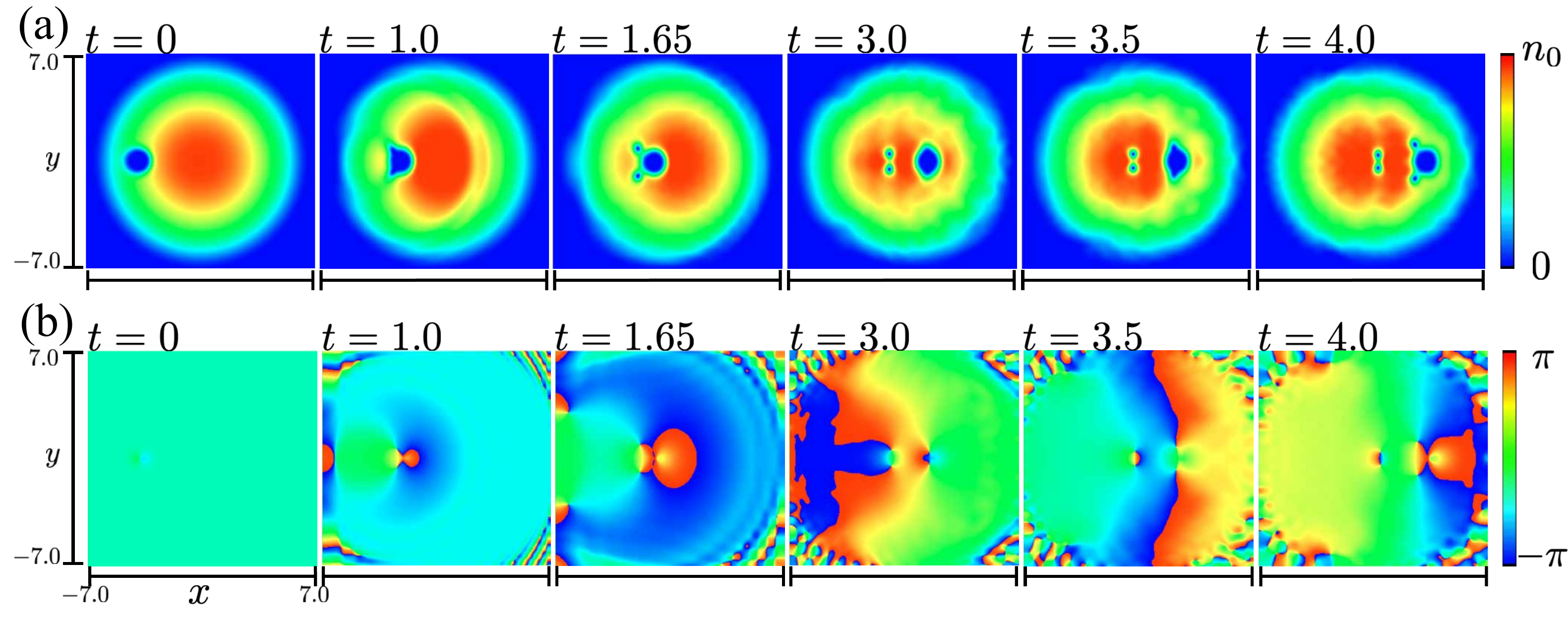}
	\centering
	\caption{The simulation result of the vortex shedding for $\mu/ (\hbar \omega) =20$, $\sigma/\xi = 4.0$, $V_0=7\mu$ and $v_{\rm obst}=0.4c_s$. 
    The panels (a) and (b) show the time evolution of the condensate density $n(\bm{r},t) = |\Psi(\bm{r},t)|^2$ and phase $\theta(\bm{r},t) = \text{arg} [\Psi (\bm{r},t) ] $, respectively. 
     The spatial region of the plot is $-7a_h \leq x,y\leq 7a_h$ .}
	\label{fig_dens}
\end{figure}

Figures \ref{fig_dens}(a) and (b) show snapshots of the condensate density $n(\bm{r},t)$ and the phase $\theta(\bm{r},t)$ for $\mu/ (\hbar \omega) = 20$, $\sigma/\xi = 4.0$, $V_0/\mu=7$ and $v_{\rm obst}=0.4c_s$, which is expected to be above the critical velocity.
When the obstacle begins to move, the obstacle pushes forward through the area of high density in the central region, inducing the collective dipole motion of the entire system as well as the small-amplitude density wave. 
Subsequently, vortex nucleation occurs at the lateral sides of the moving obstacle. 
From the phase profile in Fig.~\ref{fig_dens}(b), this is a vortex-antivortex pair created initially in the density depleted region caused by the obstacle potential.
This is slightly different from the result reported in Ref.~\cite{Kwak2023-la}, where two pairs of vortices appear initially inside the density depleted region in the impenetrable regime. 
This difference is due to the fact that our obstacle size is rather small $\sigma/\xi = 7$ compared to $\sigma/\xi = 20$ in Ref.~\cite{Kwak2023-la}. 
After the vortex pair is released from the obstacle boundary to the finite-density region, it moves to follow the obstacle's motion. 
Again, a new vortex pair is generated inside the density depleted region and released before the obstacle pass through the Thomas-Fermi boundary.
In our simulations, emission of vortex pairs occurs up to twice restricted by the finite system size.
It is noticeable that vortex nucleation and emission events tend to take place in the low-density region instead of the central high-density region.

\subsection{Method of determining the critical velocity} \label{meth}
Our finite size system has some problems for the determination of the critical velocity.
Here, we present our strategy to determine the critical velocity according to the emission frequency of quantized vortices from an obstacle \cite{limVortexsheddingfrequency2022}.

\begin{figure}[ht]
    \includegraphics[width=0.9\linewidth]{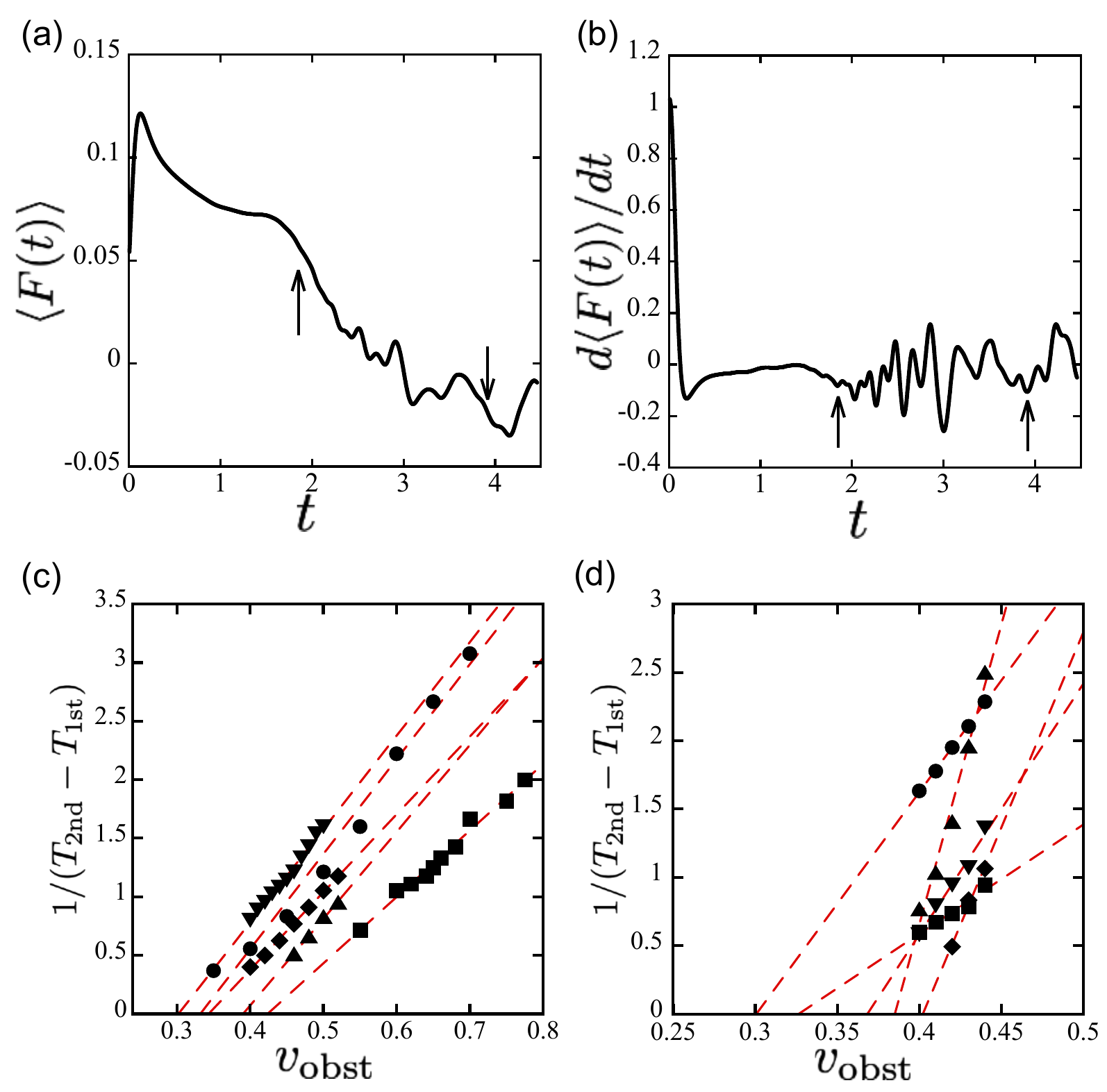}
    \centering
    \caption{ (a) The time evolution of the expected value of the drag force in the $x$-direction acting on the obstacle.
    It is difficult to get the definite time of the vortex emission from this data, where the arrows indicate the timing of the vortex emission.
    (b) The time derivative of the drag force, which enhances the subtle variation of the drag force.
    The timing of the vortex emission is identified through the comparison with the animation of the dynamics.
    The other peaks may be caused by the density waves reflected from the Thomas-Fermi boundary.
    The panels (c) and (d) show the relations between velocity of the obstacle $ v_{\rm obst}$ and the interval of the vortex emissions for $\mu/ (\hbar \omega) = 20$ and $50$.
    The plots in (c) show the results for $\sigma/\xi=2 (\blacksquare)$, $\sigma/\xi=3 (\blacktriangle)$, $\sigma/\xi=4 (\blacklozenge)$, $\sigma/\xi=5 (\bullet)$ and $\sigma/\xi=6 (\blacktriangledown)$.
    The plots in (d) show the results for $\sigma/\xi=4 (\blacklozenge)$,$\sigma/\xi=5 (\blacktriangledown)$, $\sigma/\xi=6 (\blacktriangle)$, $\sigma/\xi=8 (\blacksquare)$ and $\sigma/\xi=10 (\bullet)$.
    The value of the critical velocity is extracted at the crossing point of the horizontal axis at $(T_\text{2nd}-T_\text{1st})^{-1} = 0$ by interpolating the linear fit of the data (red-dashed lines).}
    \label{determine_vc}
    \end{figure}

When an obstacle moves through a condensate near the critical velocity $v_{\rm obst}\gtrsim v_c$, vortex dipoles are periodically generated \cite{neelyObservationVortexDipoles2010,kwonCriticalVelocityVortex2015,kwonPeriodicsheddingvortex2015}.
In the previous studies, the phase-slip event \cite{jacksonVortexFormationDilute1998} or the rapid change in a drag force acting on an obstacle \cite{reevesIdentifyingSuperfluidReynolds2015} have been proposed as methods to measure the timing of vortex emission in the numerical simulations.
In our case, since the density inside the hard potential is almost vanished, the phase fluctuation arises easily in such an extremely low-density region so that the phase slip cannot be identified clearly.
Therefore, we determine the time of vortex emissions from the drag force in the $x$-direction caused by the obstacle:
\begin{equation}
\la F(t)\ra=-\int dxdy \left(\frac{\partial}{\partial x}V_{\rm obst}\right)|\Psi|^2 .
\label{exp_ene}
\end{equation}
Figure \ref{determine_vc}(a) shows the time development of the expected value of the drag force of Eq.\eqref{exp_ene}.
Contrary to the calculations in homogeneous systems \cite{reevesIdentifyingSuperfluidReynolds2015}, it is difficult to obtain the exact time of vortex creation and emission, since the clear extremes relevant to events of the vortex emission cannot be identified. 
This is because the signal of the vortex emission is buried by the inevitable phonon emission and the density compression. To emphasize the fine changes of the time evolution of $\langle F(t) \rangle$, we take the time derivative $d \langle F(t) \rangle / dt$. 
Together with the careful check of the snapshots of the density evolution, we identify the timing of the vortex emission from the obstacle, as shown by arrows in Fig.\ref{determine_vc}(b). 

A relationship between the obstacle's velocity and the period of vortex emission $T$ is empirically given by
\begin{equation}
v_{\rm obst}-v_{\rm c} \propto \frac{1}{T}
\end{equation}
for an obstacle in a penetrable regime \cite{limVortexsheddingfrequency2022}. 
In our impenetrable regime, we assume that the critical velocity can be also obtained by this relationship.
However, it is difficult to measure such a period $T$ exactly in our finite-sized system because a number of events involving vortex emission is not enough for measuring the period.
Hence, we evaluate it approximately as $T \simeq T_\text{2nd} - T_\text{1st}$ using the interval between the first and second vortex emissions under the assumption that the vortices would be generated periodically.
As seen in Fig.\ref{determine_vc}, the interval has a linear dependence with respect to $v_\text{obst}$, which enables us to estimate the critical velocity.
\subsection{Dependence on the size of the obstacle}\label{res1}
First, we discuss the dependence of the critical velocity on the size of the obstacle, which was studied in the previous works \cite{huepeScalingLawsVortical2000,kwonCriticalVelocityVortex2015}.
The experiment in Ref.~\cite{kwonCriticalVelocityVortex2015} observed the nontrivial decreasing behavior of the critical velocity with respect to the size of the obstacle.
To investigate whether this behavior is relevant to the inhomogeneity of the system, we simulate the GP equation with a height of the potential $V_0=7\mu$ and two different values of the chemical potential $\mu/ (\hbar \omega) = 20$ and $50$.
Although these values are much smaller than the estimated value $\mu \approx 140$ corresponding to the experiment of Ref.~\cite{kwonCriticalVelocityVortex2015}, it is hard to make simulations with such a large value because of the numerical cost to describe accurately both the large condensate size and the small healing length.
Nonetheless, we expect that our choice of the parameter can capture the inhomogeneous effect since the obtained critical velocity is not so sensitive to the chemical potential, as shown below.
\begin{figure}[ht]
\includegraphics[width=0.7\linewidth]{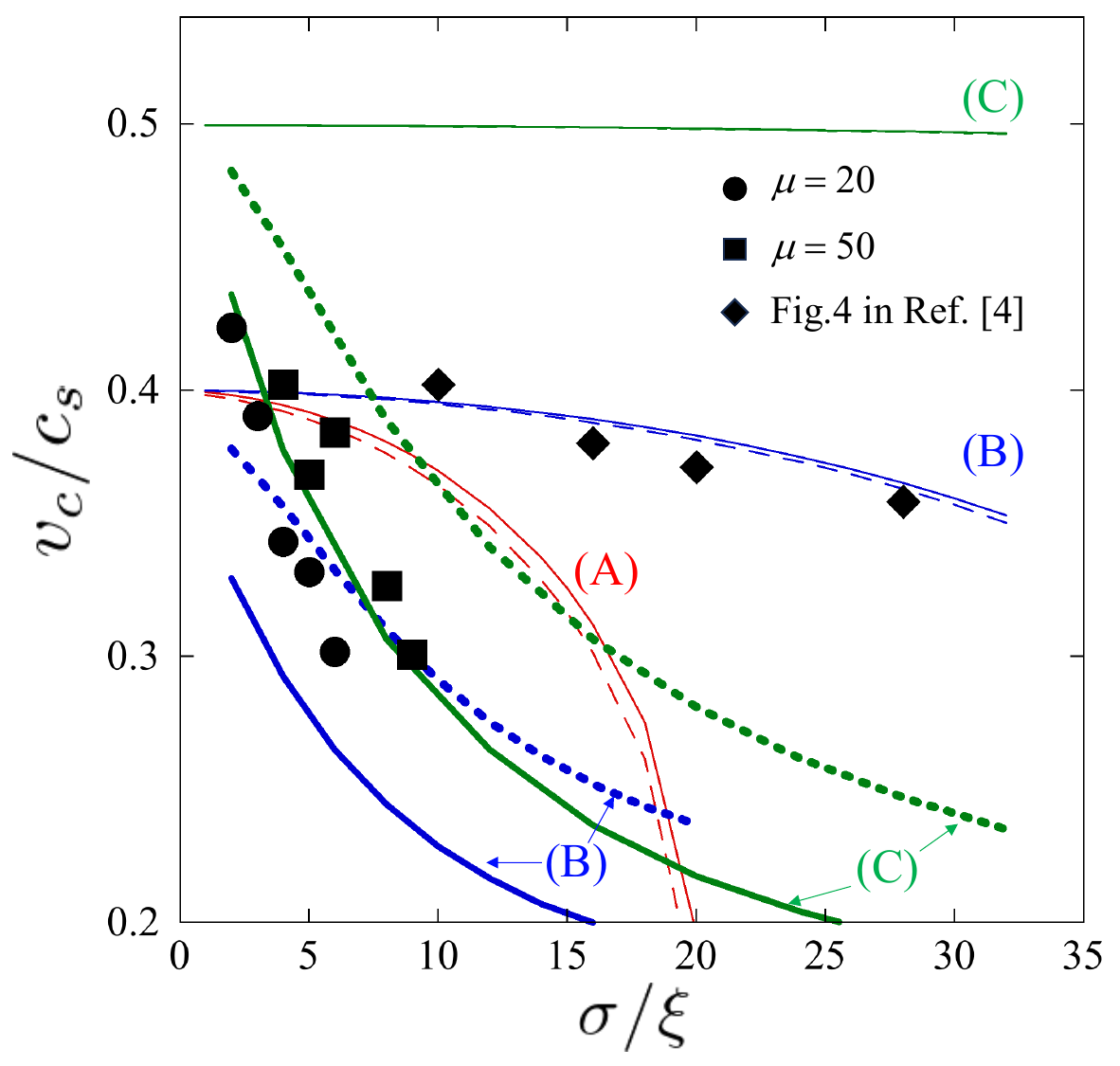}
\centering
\caption{Dependence of critical velocity on the obstacle size.
The height of the Gaussian obstacle is fixed as $V_0/\mu= 7$.
Vertical axis represents a ratio of the calculated critical velocity $v_c$ to the sound velocity $c_s$ estimated by the chemical potential at the trap center.
Filled circles and squares are results for $\mu/ (\hbar \omega) = 20$ and $50$, respectively.
Solid diamonds are experimental results taken from Fig.~4 in Ref.\cite{kwonCriticalVelocityVortex2015}.
The thin-solid and thin-dashed curves represent $v_c^\text{TF}$ of Eq.~\eqref{velohyou} for $\alpha = 1$ and 2, respectively.
Here, the choice of the parameters are (A) $(\mu,x_0) = (20\hbar\omega,-0.6R_{\rm TF})$,  (B) $(\mu,x_0) = (50\hbar\omega,-0.6R_{\rm TF})$, (C) $(\mu,x_0) = (150\hbar\omega,-0.05R_{\rm TF})$.
The Bold-solid and Bold-dotted curves represent $v_c^\text{num}$ of Eq.~\eqref{velohyou2} for $\alpha = 1$ and 2, respectively.
 }
\label{result_1}
\end{figure}

Plots in Fig.~\ref{result_1} represent the critical velocity $v_c/c_s$ of the vortex shedding, obtained by the method in Sec.~\ref{meth}, as a function of the width $\sigma$ of the Gaussian obstacle.
Although there are variations in the data, the critical velocity decreases monotonically from $\sim 0.4$ with increasing $\sigma/\xi$.
For $\sigma/\xi > 10$ the strong perturbation caused by the obstacle with the large sizes prevents the clear identification of the vortex nucleation event.
It is noticed that our results decrease more rapidly than the experimental result \cite{kwonCriticalVelocityVortex2015}.
It has been known that the critical velocity of the vortex shedding can be determined by the sound velocity multiplied by a certain numerical factor of $\lesssim \mathcal{O}(1)$.
Since a local change of the sound velocity occurs in an inhomogeneous system, the critical velocity is expected to be written by the local sound velocity $c_s^{\rm loc}(\bm{r})=\sqrt{gn(\bm{r})/m}$ under the local density approximation.
Here, we focus on the two factors that cause the density inhomogeneity.
The first one is a decrease in the density caused by the inverted-parabolic Thomas-Fermi profile.
This decrease in density should be considered seriously since the position at which the vortex nucleation occurs tends to locate far from the trap center with increasing the size of the obstacle.
The second one is the density suppression caused by the tail of the Gaussian obstacle, which becomes more remarkable as the size of the obstacle increases as seen in Fig.~\ref{grad_scale}.
The latter further brings about the decrease in the critical velocity.

It is necessary to determine the location for evaluating the local sound velocity.
Since it is difficult to determine exactly the location corresponding to the timing of the vortex emission, we consider the local density around the obstacle potential at the ``initial position” $x=x_0 (= -0.6 R_\text{TF})$.
This prescription is based on the numerical observation seen in Fig.~\ref{fig_dens} that the first vortex emission occurs near the initial position of the obstacle.
Also, we suppose that the lateral sides of the obstacle are the points where the vortex nucleation is easy to occur.
To determine this point, we consider the effective radius $R_\text{obst}$ of the obstacle, which can be estimated by solving the equation
\begin{align}
V_0 e^{-2 R_\text{obst}^2/\sigma^2} &= \mu(x_0,R_\text{obst})  \nonumber \\
&\equiv \mu - V_\text{ext}(x_0,R_\text{obst}),  \label{innerTF}
\end{align}
which corresponds to the condition determining the Thomas-Fermi boundary and is a generalization of the discussion in Ref.\cite{kwonCriticalVelocityVortex2015} to the local chemical potential. 
Thus, this condition provides the well-defined boundary induced by the Gaussian obstacle.
Equation~\eqref{innerTF} is solved numerically to obtain the value of $R_\text{obst}$, which is slightly enhanced about 10\% for $x_0=-0.6R_\text{TF}$ from the value at the trap center.
Since the vortex nucleation is likely to occur within the scale of the healing length from the boundary, the local sound velocities are determined by using the local density $n(x_0, R_\text{obst}+\alpha \xi)$. Here,the constant $\alpha \sim \mathcal{O}(1)$ is an uncertain numerical factor of order unity, including a contribution of the vortex core structure as well as a correction of the local healing length from the bulk value.
We show the results with $\alpha=1$ and 2 in the following.

We first consider only the inhomogeneous effect caused by the Thomas-Fermi density profile.
The critical velocity, scaled by the sound velocity at the trap center, is related to the local sound velocity $c_{\rm TF}^{\rm loc}$ as
\begin{equation}
v_c^\text{TF} = 0.5 \frac{c_{\rm TF}^{\rm loc}(\bm{r})}{c_s}  = 0.5 \sqrt{\frac{n_\text{TF}(x_0, R_\text{obst}+\alpha \xi)}{n_\text{TF}(0,0)}},   \label{velohyou}
\end{equation}
where we have multiplied the factor 0.5 since the local velocity at the lateral sides of the cylindrical object becomes twice of the background velocity of the fluid.
Several curves of the thin-solid ($\alpha=1$) and the thin-dashed ($\alpha=2$) lines in Fig. \ref{result_1} show $v_c^\text{TF}$ as a function of $\sigma/\xi$ for $\mu/ (\hbar \omega) = 20$ and $50$.
For $\sigma/\xi \ll 1$ the slight decrease in the density due to the inverted parabolic profile from the center  yields $v_c^\text{TF}\simeq 0.4$ smaller than $0.5$, which is consistent to the experimental \cite{kwonCriticalVelocityVortex2015} and numerical observations. 
However, these results do not exhibit the rapidly decreasing behavior obtained in the numerical results.
Also, we plot the result for the experimentally relevant parameters $\mu/ (\hbar \omega) = 150$ and $x_0 = -0.05 R_\text{TF}$ \cite{kwonCriticalVelocityVortex2015}.
Then, the value of $v_c^\text{TF}$ is almost constant around 0.5, whose behavior is largely deviated from the experimental observation.

It is clearly seen that the density around the obstacle potential is further suppressed from the Thomas-Fermi profile, which leads to additional suppression of the critical velocity.
As a next attempt, we consider the effect of density suppression caused by the Gaussian tail, which is more pronounced as the size of the obstacle increases.
To this end, we employ the local density $n_{\rm num}(x_0, R_\text{obst}+\alpha \xi)$ taken from the numerical solutions of the GP equation, as indicated in Fig.~\ref{grad_scale} for example, to evaluate the local sound velocity as
\begin{equation}
v_c^\text{num} = 0.5 \frac{c_{\rm num}^{\rm loc}(\bm{r})}{c_s}  = 0.5 \sqrt{\frac{n_\text{num}(x_0, R_\text{obst}+\alpha \xi)}{n_\text{num}(0,0)}}.   \label{velohyou2}
\end{equation}
The results are shown by the curves of bold-line ($\alpha=1$) and dotted-line ($\alpha=2$) in Fig.\ref{result_1}(b) and (c).
We find that the obtained curves for the particular value of $x_0$ are almost independent of the values of $\mu$, thus showing the results only for $\mu/ (\hbar \omega) = 50$ in the case of $x_0 = - 0.6R_{\rm TF}$.
The result indicates that including the density suppression by the Gaussian tail can reproduce the rapid decreasing behavior of the critical velocity with respect to $\sigma$.
We see that the numerical results are fitted better for $\alpha=2$.
We also find that the decreasing behavior obeys the power low $\sim (\sigma/\xi)^{1/3}$, which is a decay slower than the predictions of the previous studies \cite{Kwak2023-la,stiessbergerCritcalVelocitySuperfluid2000}.
Furthermore, we plot the case of the experimentally relevant parameters ($\mu/ (\hbar \omega) = 150$ and $x_0=-0.05R_{\rm TF}$).
Since the initial position of the Gaussian obstacle is near the center of the condensate, the estimated critical velocity is larger than that for $\mu/ (\hbar \omega) = 50$.
However, the values underestimate the experimental results.

We give two comments before enclosing this section.
\begin{enumerate}
\item The value of $v_c^\text{num}$ does not depend on the chemical potential $\mu$, while the numerical results indicate that $v_c$ for $\mu/ (\hbar \omega) = 20$ is clearly smaller than that for $\mu/ (\hbar \omega) = 50$.
One possible reason of this discrepancy may be considered as follows.
The flow induced by the moving obstacle may results in the backflow due to the reflection of the density wave from the outer Thomas-Fermi boundary.
This backflow enhances the relative velocity between the condensate and the obstacle, resulting in the decrease in $v_c$.
For a small condensate with $\mu/ (\hbar \omega) = 20$ and a fast moving obstacle, this effect cannot be negligible.
\item Our result seems not to be consistent with the experimental observation in Ref.~\cite{kwonCriticalVelocityVortex2015}.
In the experiment \cite{kwonCriticalVelocityVortex2015}, the critical velocity was measured by seeing a single event whether the vortices appear or not through the displacement $\Delta x = 0.1 R_{\rm TF}$, which seems to be small compared with the whole condensate size.
As shown in the next section, this experimental protocol may lead to overestimation of the critical velocity compared with that with the method in Sec.~\ref{meth} using the period of the vortex emission.
\end{enumerate}

\subsection{Dependence on the nonlinear coefficient}\label{res2}
\begin{figure}[ht]
	\includegraphics[width=0.8\linewidth]{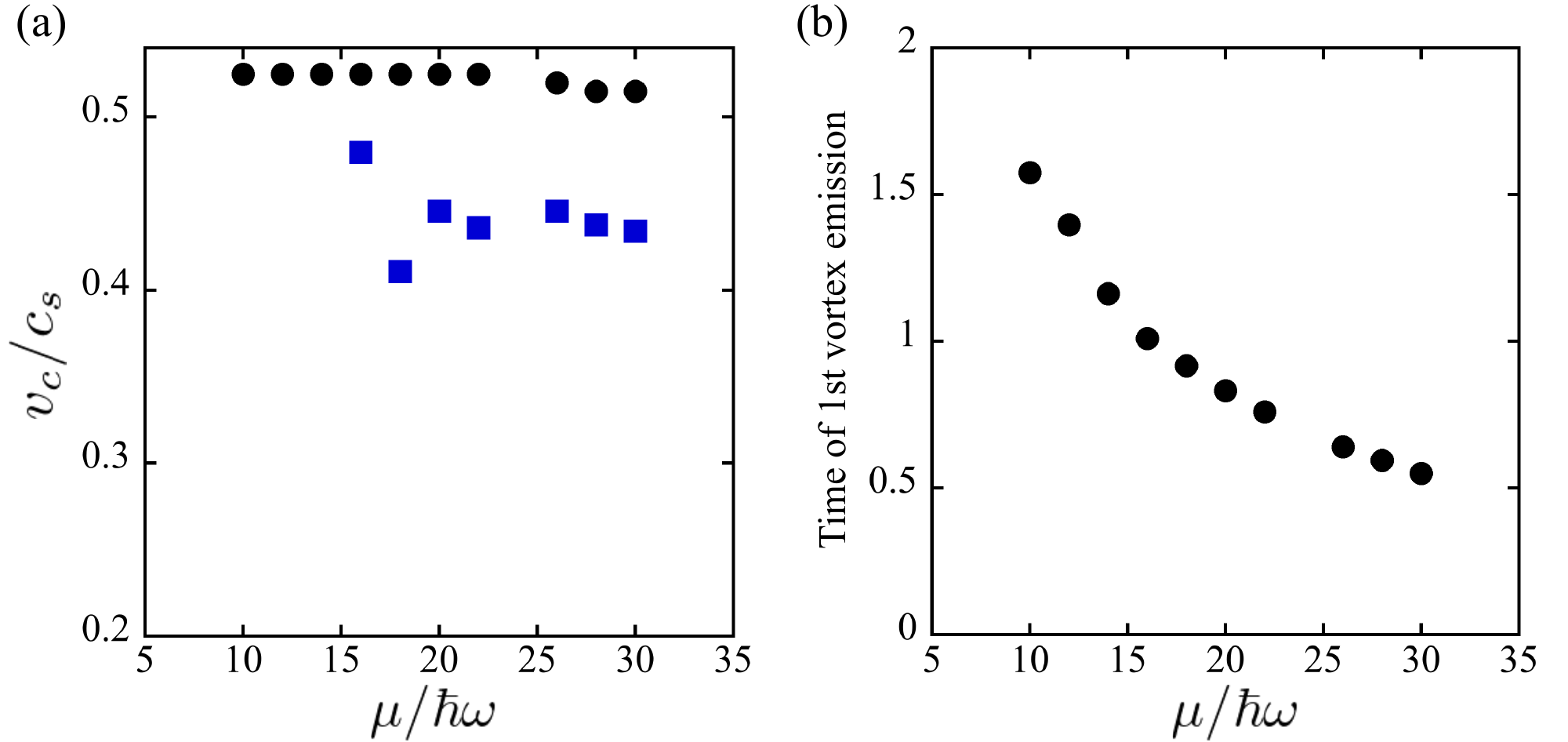}
	\centering
	\caption{Dependence of critical velocity on the chemical potential $\mu$ for $V_0=10\mu$ and $\sigma/\xi=0.6\sqrt{5}\sim1.34$.
	In (a), the vertical axis is a ratio of the critical velocity $v_c$ to the sound velocity $c_s$ at the center of the system.
        The squares are obtained by the method in Sec.\ref{meth}, while the circles are obtained by seeing whether the vortex emission occurs in the simulations. 
        (b) The time duration of the first event of the vortex emission as a function of $\mu$.}
	\label{result_2}
\end{figure}

As shown in the previous section, the critical velocity, scaled by the sound velocity at the center, is insensitive to the chemical potential i.e., the nonlinear coefficient of the GP equation, even in the trapped system.
Here, as supplemental results in the previous section, we investigate the dependence of the critical velocity on the nonlinear coefficient of the GP equation.
We fix $V_0=10 \mu$ and the obstacle size $\sigma/\xi=0.6\sqrt{5}\sim1.34$ as small as possible.
The latter choice is to reduce the effect of the Gaussian inhomogeneity described in Sec.\ref{res1}.
Figure \ref{result_2}(a) shows the critical velocity as a function of $\mu$.
Here, in addition to the method in Sec.~\ref{meth}, we give an alternative estimate of the critical velocity by seeing whether the event of the vortex emission from the obstacle occurs.
In fact, the latter method follows the protocol used in the experiment of Ref.\cite{kwonCriticalVelocityVortex2015}.
The two methods confirm that the critical velocity is insensitive to the value of $\mu$.
These results are evident from the density profile of the initial state in Fig.\ref{grad_scale}; the local density $n_\text{num}(x_0,R_\text{obst} + \alpha \xi)$ decreases as $\sigma$ increased for the fixed $\mu$, while it is almost unchanged by varying $\mu$ for the fixed $\sigma$.
The critical velocity obtained by the latter method is always larger than the former, since this method does not eliminate the possibility of vortex generation in the successive time evolution.
This might be a possible reason why the critical velocities in the experiment \cite{kwonCriticalVelocityVortex2015} are always larger than our numerical results.  

The method in Sec.~\ref{meth} is not applicable for small $\mu \lesssim 15$ since the time duration for which the vortex nucleation occurs is extended, as shown in Fig.~\ref{result_2}(b).
Also, the result in Fig.~\ref{result_2}(b) implies that vortex nucleation is less likely to occur as the chemical potential decreases.
In a uniform system, the dynamics is free from the chemical potential since there are no parameters in the scaled GP equation. Thus, the tendency observed in Fig.\ref{result_2}(b) can be interpreted as an inhomogeneous effect, indicating that the vortex nucleation is not easy to occur for small $\mu$ since the healing length is extended over the system so that the local vortex nucleation may be prohibited.

\section{Conclusion}\label{sec4}
In this study, we investigated the influence of inhomogeneity of the trapped Bose-condensed system on the critical velocity for vortex shedding via a Gaussian obstacle. The decrease in the critical velocity from one estimated in a homogeneous bulk sound velocity can be explained by the local sound velocity due to not only the inverted parabolic density profile in a trap potential but also the density suppression by the tail of the Gaussian potential. We also confirmed that the critical velocity is not dependent on the value of the chemical potential, which indicate that universal discussion in a uniform system is approximately applicable to the trapped system by properly including the inhomogeneous feature within the local density approximation.  

\bibliography{sn-ref}

\end{document}